\documentclass{emulateapj}

\shorttitle{Fundamental Parameters of NGC1245}
\shortauthors{Burke et~al.}

\begin{document}

\title{Survey for Transiting Extrasolar Planets in Stellar Systems:\\
I. Fundamental Parameters of the Open Cluster NGC 1245}
\author{Christopher J. Burke}
\affil{Astronomy Department, The Ohio State University}
\affil{140 W. 18th Ave., Columbus, OH 43210}
\email{cjburke@astronomy.ohio-state.edu}
\author{B.\ Scott Gaudi}
\affil{Harvard-Smithsonian Center for Astrophysics}
\affil{60 Garden Street, Cambridge, MA 02138}
\email{sgaudi@cfa.harvard.edu}
\and
\author{D. L. DePoy, Richard W. Pogge, Marc H. Pinsonneault}
\affil{Astronomy Department, The Ohio State University}
\affil{140 W. 18th Ave., Columbus, OH 43210}
\email{depoy, pogge, pinsono@astronomy.ohio-state.edu}

\begin{abstract} We derive fundamental parameters for the old, rich
open cluster NGC 1245 using $BVI$ photometry from the MDM 1.3m and
2.4m telescopes.  Based on detailed isochrone fitting, we find NGC
1245 has a slightly sub-solar metallicity, ${\rm [Fe/H]}=-0.05\pm
0.03~{\rm (statistical)} \pm 0.08~{\rm (systematic)}$, and an age of
$1.04\pm 0.02\pm 0.09$ Gyr.  In contrast to previous studies, we find
no evidence for significant differential reddening.  We determine an
extinction of $A_{V}=0.68\pm0.02\pm 0.09$ mag and a distance modulus of
$(m-M)_{0}=12.27\pm 0.02\pm 0.12$ mag, which corresponds to a distance of
$2.8 \pm 0.2~{\rm Kpc}$.  We derive a logarithmic mass-function slope
for the cluster of $\alpha=-3.12\pm 0.27$, where a Salpeter slope is
$\alpha=-1.35$.  Fits to the radial surface-density profile yield a
core radius of $r_c=3.10\pm 0.52~{\rm arcmin}$ $(2.57\pm 0.47~{\rm
pc})$.  NGC 1245 is highly relaxed and contains a strongly mass
segregated population.  The mass function for the inner cluster has a
very shallow slope, $b=-0.56\pm\,0.28$.  In contrast, the outer
periphery of the cluster is enriched with low mass members and devoid
of high mass members out to the tidal radius, $r_{t}=20$ arcmin
(16.5~${\rm pc}$).  Based on the observed surface-density profile and
an extrapolated mass function, we derive a total cluster mass,
$M=1300\pm 90\pm 170 M_{\odot}$.  \end{abstract}


\section{Introduction}

Open clusters are excellent laboratories for many different aspects of
astrophysics.  First, open clusters form a coeval set of stars with
homogeneous properties.  This homogeneity makes it possible to
determine with relative ease the age and metallicity of the cluster
\citep{YI01}.  Second, owing to their relatively small relaxation
times, open clusters provide an opportunity to study stellar systems
in various stages of dynamical evolution \citep{BIN87}.  Finally,
theory predicts the star formation process within an open cluster is
strongly affected by dynamical interactions, supernovae explosions,
and UV radiation \citep{ADA01}.  Characterizing the fraction of
cluster members with companions from stellar to planetary masses for a
cluster in comparison to the field provides valuable insight into how
these processes affect stellar and planetary formation.

The large angular size (tens of arcminutes to several degrees) of open
clusters has traditionally made it difficult and time consuming to
observe a substantial fraction of an open cluster's members.  However,
recently available large-format CCD imagers now allow for complete
studies of an open cluster with relatively small amounts of observing
time.  In this paper we present wide-field $BVI$ photometry of the
open cluster NGC 1245 using the MDM 8K Mosaic imager on the MDM 2.4m
Hiltner telescope and supplemental $BVI$ photometry obtained under
photometric conditions on the MDM 1.3m McGraw-Hill telescope.

NGC 1245 is a rich, old open cluster with an approximately solar
metallicity population \citep{JAN88, WEE96}.  The cluster is located
toward the galactic anticenter ($\ell=147^{\circ}$, $b=-9^{\circ}$) at
a distance, $R\sim2.5$ kpc \citep{JAN88, WEE96}.  With its relatively
large Galactocentric distance, NGC 1245 is particularly useful in
constraining the Galactic metallicity gradient \citep{FRI95,WEE96}.
Here, we improve on the observations for this cluster by covering a
six-times greater area, obtaining the first CCD $I$-band photometric
data, and acquiring the first CCD data under photometric conditions.
With these significant improvements over earlier studies, we are able
to determine more precisely the physical parameters of the cluster as
a whole, rather than just the inner cluster core.  In addition to the
observational improvements, we improve the interpretation of the
observations by calculating the physical parameters of NGC 1245 using
a chi-square fit to the observations, and we make every effort to
quantify the systematic errors in the derived physical parameters.

Because NGC 1245 is a rich open cluster and readily observable from
our primary observing site, it is the first target of the Search for
Transiting Extrasolar Planets in Stellar Systems (STEPSS) project
\citep{BUR03}.  The goal of the STEPSS project is to assess the
frequency of close-in extrasolar planets around main sequence stars in
several open clusters.  Part of the $BVI$ data we are presenting in
this paper were taken during a 19 night observing run to search for
transiting extrasolar planets in NGC 1245.  This transit data consist
of $I$-band only observations obtained every seven minutes with the
MDM 8K Mosaic imager on the MDM 2.4m telescope.  Results of the search
for extrasolar planets in NGC 1245 using this transit data will be
presented in an upcoming paper (B.\ S.\ Gaudi et~al., in
preparation).  An additional paper presenting the variable star
content of our NGC 1245 field based on light curves from the transit
data is also in progress (J. Pepper et~al., in preparation).  The
main goal of this paper is to determine the fundamental
physical parameters of NGC 1245.  With the known properties for this
cluster and subsequent clusters in the STEPSS project, we will gain
insight into how metallicity, age, and stellar density affect
planetary formation, migration, and survival.

We describe our observations and data reduction in \S \ref{OBS}.  In
\S\ref{PARS}, we derive the physical parameters of NGC 1245 from
isochrone fits to the $BVI$ photometry.  We derive the radial profile
of NGC 1245 in \S\ref{RAD}, the mass function in \S\ref{massfunc}, and
the total mass in \S\ref{totmass}.  We summarize and conclude in
\S\ref{CON}.

\section{Observations and Data Reduction}\label{OBS}

\subsection{Observations}

\begin{figure}
\plotone{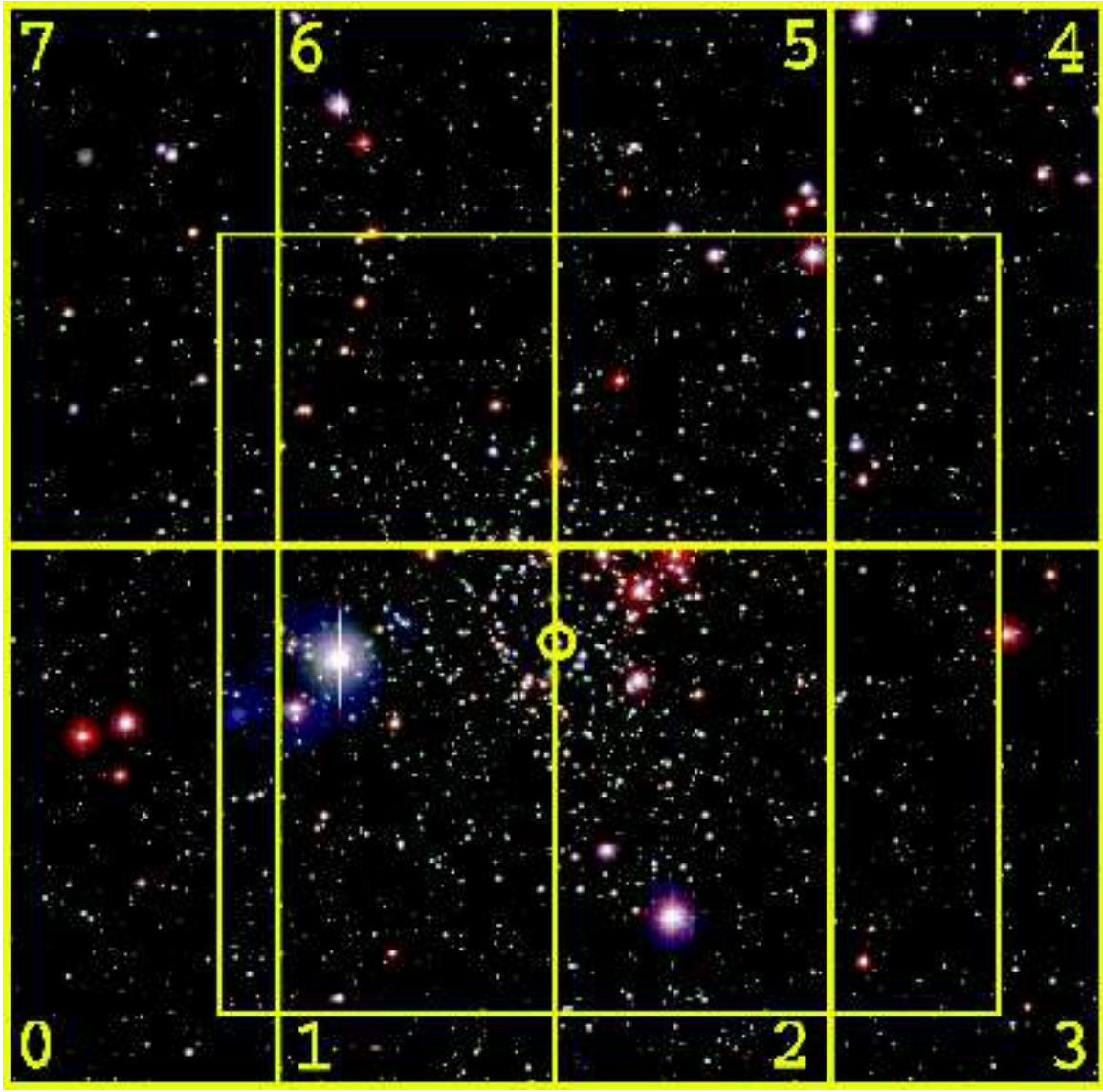}
\caption{A false color image of the open cluster NGC 1245 obtained
with the MDM 8K Mosaic imager (the 4x2 array of CCDs are numbered) on
the MDM 2.4m telescope.  The large inscribed square is the field of view 
for the MDM 1.3m telescope data obtained with a single CCD
imager.  The blue, green, and red color channels of the false color
image consist of combined images in the $BVI$ passbands, respectively.
The circle shows the cluster center.}
\label{n1245fov}
\end{figure}

We observed NGC 1245 on two occasions.  The first set of observations
was obtained in November 2001 using the MDM 8K mosaic imager on the
MDM 2.4m Hiltner telescope.  The MDM 8K imager consists of a 4x2 array
of thinned 2048x4096 SITe ST002A CCDs \citep{CRO01}.  This
instrumental setup yields a 26$\arcmin$x26$\arcmin$ field of view and
a 0.36$\arcsec$ per pixel resolution employing the 2x2 pixel binning
mode.  Table~\ref{obsdat24} shows for each night of observations the
number of exposures in each filter, exposure time, median full width
at half maximum (FWHM), and a brief comment on the observing
conditions.  This observing run is optimized for a transiting
extrasolar planet search, thus there are 960 I-band images obtained
during this run that are not listed in Table~\ref{obsdat24}.  Since
the emphasis of the present study is the color-magnitude diagram of NGC 1245,
we chose only the I-band images on each night that were taken
concurrent with the other two filters and the same number of images to
have similar signal to noise characteristics for each filter.  None of
the nights were photometric.  Therefore, we reobserved NGC 1245 in
February 2002 using the MDM 1.3m McGraw-Hill telescope with the
2048x2048 ``Echelle'' imaging CCD.  This CCD provides a
17$\arcmin$x17$\arcmin$ field of view with 0.5$\arcsec$ per pixel
resolution.  Several nights were photometric, and we use Landolt
\citep{LAN92} standard star observations to calibrate our photometry.
Table~\ref{obsdat13} details the images taken with the MDM 1.3m
telescope.  The relative placement between the 1.3m and 2.4m field of
views is shown in Figure~\ref{n1245fov}.

Figure~\ref{n1245fov} shows a false-color image of NGC 1245 and shows
the overlap between the MDM 2.4m and 1.3m datasets.  The grid of eight
numbered rectangles shows the field of view for the eight CCDs that
make up the MDM 8K Mosaic imager.  The large inscribed square is the
field of view of the MDM 1.3m observations.  The blue, green, and red
color channels for the false-color image result from a median combined
image of the $BVI$ passbands, respectively.  The circle denotes the
cluster center derived in Section~\ref{RAD}.

\subsection{Data Reduction}

We use the IRAF\footnote{IRAF is distributed by the National Optical
Astronomy Observatories, which are operated by the Association of
Universities for Research in Astronomy, Inc., under cooperative
agreement with the National Science Foundation.}  CCDPROC task for all
CCD processing.  We describe the 2.4m data reduction first.  The
stability of the zero-second image over the course of the 19 nights
allows us to median combine 95 images to determine a master
zero-second calibration image.  For master flat fields, we median
combine 11, 19, and 66 twilight sky flats taken throughout the
observing run in the $BVI$ passbands, respectively.  We quantify the
errors in the master flat field by examining the night to night
variability between individual flat fields.  The small-scale,
pixel-to-pixel variations in the master flat fields are $\sim 1\%$,
and the large-scale, illumination-pattern variations reach the 3\%
level.  The large illumination-pattern error results from a strong
sensitivity in the illumination pattern to telescope focus.  However,
such large-scale variations do not affect photometry where the
point-spread-function (PSF) scale is the most relevant.

We use the DAOPHOT package \citep{STE87} within IRAF for PSF fitting
photometry, and we perform the photometry on each frame individually.
To identify stars and to calculate stellar positions, we designate the
highest signal-to-noise $I$-band image as input to the DAOFIND task.
We transform the stellar positions determined on the high
signal-to-noise $I$-band image to all the other images as initial
guesses for photometry.  To calculate the model PSF, we select 25
bright, isolated stars evenly distributed on the reference image.  We
use the same PSF stars for all images.  Slight offsets between images
result in at least twenty stars in the PSF model calculation for all
the images.  Modeling the spatially constant PSF model involves three
iterations with faint neighbors to the PSF stars being subtracted from
the image between iterations.  Using the resulting PSF model, the IRAF
ALLSTAR task calculates the instrumental magnitudes by fitting the PSF
model to all stars simultaneously.  The reference image photometry
defines our instrumental magnitude system, and we apply a zero-point
offset to the individual image photometry to align them with the
reference image photometry.  A robust weighted mean using the errors
output by ALLSTAR gives the final instrumental magnitude.

\begin{figure}
\plotone{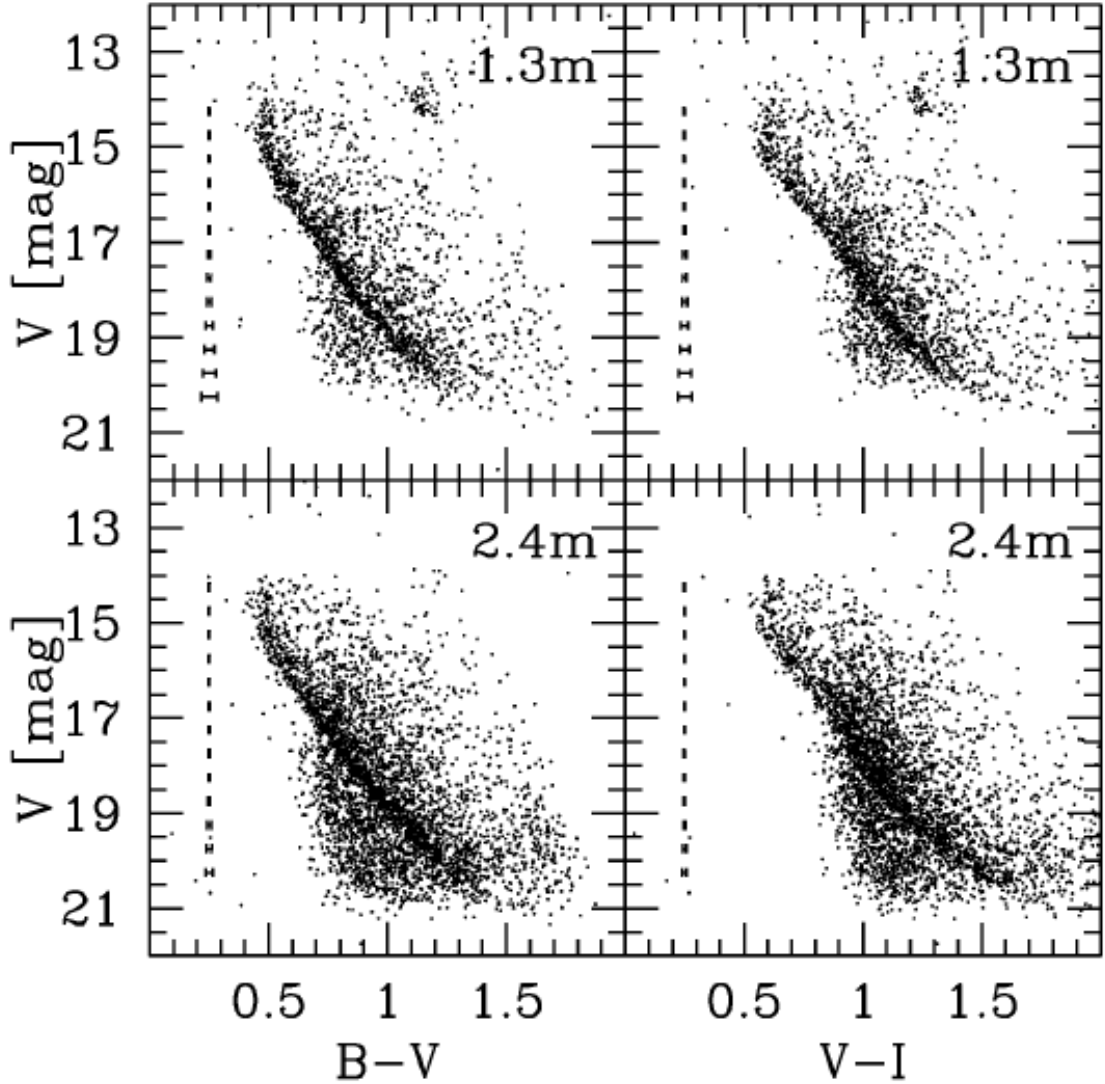}
\caption{ Color magnitude diagrams (CMDs) of the open cluster NGC
  1245.  Error bars represent the median sample standard deviation in
  the stellar color as a function of magnitude.  Top panels: Data from
  the MDM 1.3m.  Bottom panels: Data from the MDM 2.4m.  Left panels:
  $V$ versus $(B-V)$.  Right panels: $V$ versus $(V-I)$.}
\label{cmds}
\end{figure}

In order to place the MDM 2.4m instrumental magnitudes onto a standard
system, we use the IRAF FITPARAMS task to determine the transformation
to the photometric MDM 1.3m dataset.  The rms scatter for an
individual star around the best fit photometric transformation is
$\sim$0.03 mag in all passbands.  This error represents how well an
individual star approximates the photometric system defined by the
1.3m data described below.  The data taken on CCDs 1 and 3 of the MDM
8K mosaic imager suffered from a nonlinearity that reaches a level of
0.14 mag near full well.  With the aid of the 1.3m dataset, this
nonlinearity is correctable such that the errors in the photometric
calibration for these CCDs are similar to the other CCDs not affected
by the nonlinearity.  The bottom panels of Figure~\ref{cmds} show the
resulting color magnitude diagrams (CMDs) for the MDM 2.4m dataset.
The stars shown have sample standard deviations in the $BVI$
photometry of less than 0.08 mag and at least 3 (B), 3 (V), and 5 (I)
photometric measurements return from the photometry package without
numerical convergence errors.  The error bars in Figure~\ref{cmds}
show the median of the sample standard deviations in the photometric
color determination for stars in bins of 0.5 mag at V.  These error
bars represent the expected deviation from the mean of a color
measurement from a single set of $BVI$ exposures with similar signal
to noise properties as our dataset.  Our ability to determine the mean
color for a star requires dividing these errors by the $\sqrt{N}$ of
the number of measurements.

\begin{figure}
\plotone{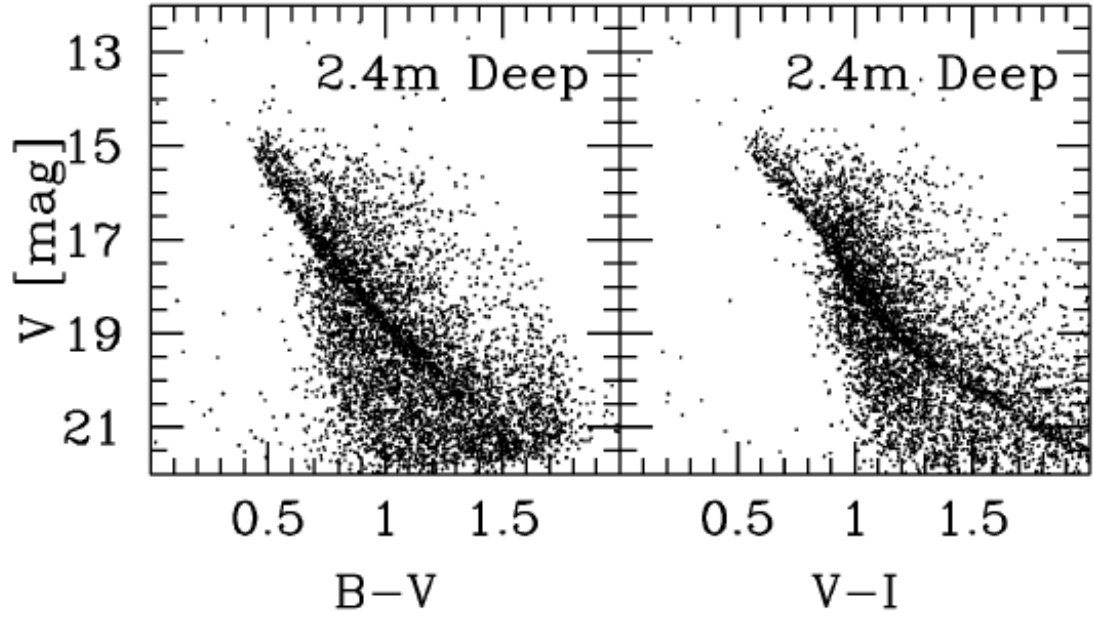}
\caption{Deep CMDs of the open cluster NGC 1245 calculated by median
combining the MDM 2.4m data.  Left panel: $V$ versus $(B-V)$.  Right
panel: $V$ versus $(V-I)$.}
\label{cmds2}
\end{figure}

An alternative method to analyze the MDM 2.4m dataset is to combine
the images and perform the photometry on this master combined image.
The combined image enhances the signal to noise of the faintest stars.
This enhanced signal to noise allows DAOFIND to detect and ALLSTAR to
fit the model PSF for significantly fainter stars at the expense of
having to trust the error output from the photometry package.  We
median combine 15, 30, and 31 $BVI$ images, respectively, and perform
the identical photometry procedure outlined above.  Initial guesses
for the stellar positions come from the combined I-band image.  The
instrumental magnitudes are fit to the calibrated MDM 1.3m dataset
with a simple zeropoint and color term.  The resulting CMDs are shown
in Figure~\ref{cmds2}.  As one can see these CMDs go significantly
fainter, but the field star contamination becomes increasingly a
problem, especially for the $V,B-V$ CMD.  The low contrast of the
cluster's main sequence against the field star contamination limits
the utility of these deeper CMDs in determining the physical
parameters of NGC 1245.  In this study we only use these deep CMDs for
extending our mass function toward fainter magnitudes (see
Section~\ref{massfunc}).

The analysis procedure for the MDM 1.3m dataset is similar to the MDM
2.4m dataset.  The master zero-second image consists of a median
combination of 49 images, and a master sky flat consists of a median
combination of 12, 15, and 4 images for $BVI$ passbands, respectively.
The PSF has strong spatial variations, so we choose the same set of
$\sim 80$ stars evenly distributed across a reference image to
calculate a quadratically varying PSF model for each image
individually.  We employ the stand-alone ALLFRAME program
\citep{STE98} to calculate the instrumental magnitudes for all images
simultaneously.  The aperture corrections also are spatially variable.
We fit a third order 2-D polynomial to the aperture corrections
calculated on the evenly distributed PSF stars.  The IRAF XYZTOIM task
provides the best fit solution for the 2-D polynomial aperture
correction model.  The standard deviation of the aperture correction
is $\sim 0.02$ mag.

Aperture photometry measurements of the Landolt field SA101
\citep{LAN92} taken at three different airmasses, $X=$1.2, 1.5, and
1.9, on two separate photometric nights (see Table~\ref{obsdat13})
provide a determination of the airmass coefficient.  The airmass
coefficients for the two nights are statistically identical; thus, we
average both airmass coefficient determinations for the final
calibration.  Fixing the airmass coefficients, we calculate color
terms for each night using observations of the Landolt fields G44-27,
G163-50/51, and PG1407 \citep{LAN92}.  The color coefficients from the
individual nights are also statistically identical, and we use the
mean of both color coefficient determinations for the final
calibration. The standard deviations in the calibrations are 0.015,
0.015, and 0.02 mag for the $BVI$ passbands, respectively.  This
represents how well an individual star approximates the photometric
system outlined by \citet{LAN92}.

There are 2, 2, and 3 images in the $BVI$ passbands, respectively, of
NGC 1245 that are photometric.  After applying the aperture
corrections to the ALLFRAME instrumental magnitudes for the
photometric data, we take a weighted mean of the measurements and then
apply the photometric calibration to derive an initial standard-star
catalog.  The final standard-star catalog is obtained by applying the
color term and aperture corrections to all nonphotometric MDM 1.3m
data, fitting for the zero-point offset to the initial standard-star
catalog, and calculating the robust weighted mean of all measurements.
  
In the process of determining the zero-point fit for the
nonphotometric MDM 1.3m data to the initial standard-star catalog, we
uncovered a 1.5\% amplitude residual that correlates with spatial
position on the CCD.  The residual is negligible over most of the CCD,
but the residual abruptly appears for stars with a position $x\la500$
in pixel coordinates.  The residual linearly grows toward decreasing
$x$ position on the detector reaching the 1.5\% level at the edge.
Due to the abrupt onset of this residual, even a cubicly varying PSF
model did not reduce the amplitude of the residual.  To reduce the
effect of this systematic spatial variability, we limit the zero-point
fit for the nonphotometric MDM 1.3m data to stars in the middle of the
detector defined by $500<x<1500$ and $500<y<1500$ in pixel
coordinates.

The top panels in Figure~\ref{cmds} show the resulting CMDs for the
MDM 1.3m dataset.  Only stars that have at least 3 photometric
measurements return from the photometry package without numerical
convergence errors and a sample standard deviation $<$ 0.08 mag in all
three passbands are shown.

\subsection{Comparison with Published Photometry}

There is some confusion in the literature regarding the photometric
calibration of NGC 1245.  \citet{CAR94} perform the first CCD detector
study of NGC 1245.  They calibrate their photometry by comparison with
the photoelectric observations of \citet{HOA61}.  Unfortunately, the
data of \citet{CAR94} available on the widely used WEBDA open cluster
database\footnote{http://obswww.unige.ch/webda/} is incorrect since it
lacks the photometric zeropoint as used in publication (G.~Carraro
private communication).  Thus, using the incorrect data from WEBDA,
\citet{WEE96} concluded the calibration of \citet{CAR94} was
incorrect.  Our independent photometric calibration agrees with the
earlier photoelectric studies.  For 5 and 3 stars in common with the
photoelectric studies of \citet{HOA61} and \citet{JEN75},
respectively, we find $V$-band differences of $0.03\pm0.06$ and
$-0.012\pm0.007$ mag, respectively, and $(B-V)$ color differences of
$-0.01\pm0.01$ and $-0.001\pm0.02$ mag, respectively.  The $(B-V)$
colors of the stars used in the comparison span the range from 0.6-1.2
mag.  Thus, the photometric calibration for NGC 1245 is well
determined.

The recent study by \citet{SUB03} does not find any differences with
the \citet{CAR94} photometry in the $V$ band but does find a linear
trend in the $(B-V)$ color difference as a function of $V$ magnitude
with an amplitude of 0.35 mag.  It is unclear if the comparison with
the \citet{CAR94} photometry in \citet{SUB03} is with the incorrect
data from WEBDA.  However, we do not find any (B-V) color difference
trends with the \citet{CAR94} data.  Additionally, inspection of the
\citet{SUB03} CMD finds the giant clump is 0.35 mag redder in $(B-V)$
than in our CMD.  The above comparisons call into question the
\citet{SUB03} color calibration.

\section{Physical Parameters}\label{PARS}
NGC 1245 is located in the Galactic plane ($b=-9^\circ$), and therefore,
the observed field contains a large number of foreground and
background disk stars complicating the process of fitting theoretical
isochrones to the CMDs.  We attempted to statistically subtract the
field-star population with a procedure similar to \citet{MIG98}.
Unfortunately, because the contrast of the main sequence above the
background field stars was relatively low (in comparison to the work
by \citet{MIG98}), we were unable to achieve a satisfactory background
field subtraction.

\begin{figure}
\plotone{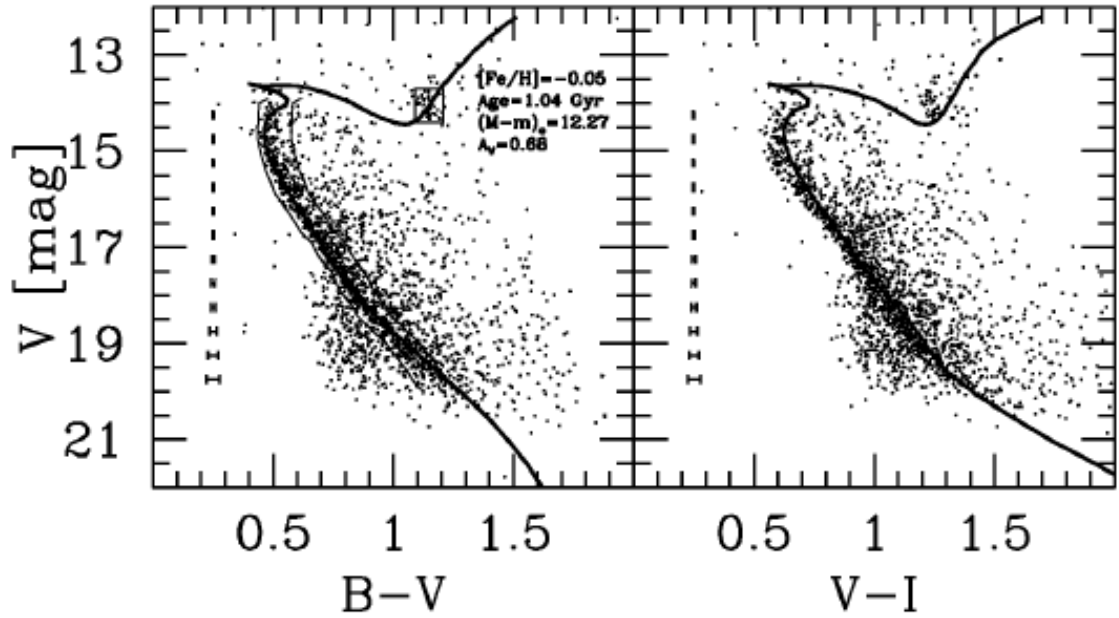}
\caption{The heavy solid line represents the best fit isochrone
solution overlying the color magnitude diagrams based on the MDM 1.3m
data.  The light solid lines show the main sequence selection
boundaries.}
\label{isofitcmd}
\end{figure}

Rather than statistically subtracting field contamination, we select
stars for the main sequence isochrone fit by the following procedure.
Since binaries and differential reddening affect the $(B-V)$ color of
a star less than its $(V-I)$ color, the $V,(B-V)$ CMD has a tighter
main sequence than the $V,(V-I)$ CMD.  The tighter main sequence in
the $V,(B-V)$ CMD has a higher contrast above the contaminating field
stars.  Thus, to select main sequence cluster members for the
isochrone fit, we trace by eye the blue main sequence edge in the
$V,(B-V)$ CMD.  Shifting this trace redward by 0.12 mag defines the
red edge selection boundary.  Finally, shifting the trace 0.02 mag
blueward defines the blue edge selection boundary.  These shifts were
selected by eye to contain a high fraction of the single-star main
sequence.  The vertical light solid lines in Figure~\ref{isofitcmd}
show the resulting main sequence selection boundaries in the $V,(B-V)$
CMD.

Since the difference in color between the red-giant clump and the main
sequence turnoff in the CMD places strong constraints on the cluster
age, we designate a box in the $V,(B-V)$ CMD to select red giant
members for the isochrone fit.  The region with boundaries
$13.7<V<14.4$ and $1.09<(B-V)<1.21$ defines the red-giant clump
selection.  The light solid box around the red-giant clump in the
$V,(B-V)$ CMD in Figure~\ref{isofitcmd} shows the selection criteria
for these stars.  The red-giant clump is saturated in the MDM 2.4m
dataset, and therefore, we fit isochrones to the MDM 1.3m photometry
only.  There are 1004 stars that meet our main sequence and red-giant
clump selection criteria.

For the isochrone fitting, we use the Yale-Yonsei (Y$^{2}$) isochrones
\citep{YI01}, which employ the \citet{LEJ98} color calibration.  The
Y$^{2}$ isochrones provide an interpolation scheme to calculate
isochrones for an arbitrary age and metallicity within their grid of
calculations.  For a given set of isochrone parameters (metallicity,
age, distance modulus, $A_{V}$, and $R_{V}$), we define the goodness
of fit as 
\begin{equation} \chi^2_{tot}=\sum_{i}\chi^2_i,
\label{chitot}
\end{equation}
where the sum is over all stars selected
for isochrone fitting, and $\chi^2_i$ is the contribution from star
$i$,
\begin{equation} \chi^2_i\equiv
(B_{pred}(m_{i})-B_{obs,i})^{2}+(V_{pred}(m_{i})-V_{obs,i})^{2}+(I_{pred}(m_{i})-I_{obs,i})^{2}.
\label{isochi} 
\end{equation}
In the preceding equation, $BVI_{pred}$
are the magnitudes predicted by the isochrone, using the stellar mass,
$m_{i}$, as the independent variable, and $BVI_{obs,i}$ are the
observed stellar magnitudes.  We use the Brent minimization routine
\citep{PRE01} to determine the stellar mass that minimizes $\chi^2_i$
for each star.

Theoretical isochrones and color calibrations contain systematic
uncertainties that make it impossible to find a consistent fit to CMDs
of the best studied open clusters over a wide range of colors and
magnitudes \citep{DEB01,GRO03}.  In order to avoid overemphasizing the
main sequence turnoff, where the photometric errors are the smallest,
we adopt an equal weighting to all stars.

To find the best fit isochrone parameters, we perform a grid search
over age and metallicity.  At fixed age and metallicity, we use the
Powell multidimensional minimization routine \citep{PRE01} to
determine the best fit distance modulus and reddening, $A_{V}$, by
minimizing $\chi^2_{tot}$.  To determine the reddening in the other
passbands, $A_{B}$ and $A_{I}$, we must assume a value for $R_{V}$,
the ratio of total to selective extinction \citep{CAR89}.  For our
best fit solution, we fix $R_{V}=3.2$.  We discuss the reasoning for
this choice and its impact on the results in the
Section~\ref{syserror}.  We calculate $\chi^2_{tot}$ over a 40x40
evenly-log-spaced grid of metallicity and age that ranges $-0.26\leq
{\rm [Fe/H]} \leq 0.13$, and $0.7943\leq {\rm Age (Gyr)}\leq 1.4125$.

We determine our confidence limits on the model parameters by scaling
the resulting chi-square statistic,
\begin{equation}
\Delta\chi^{2}=\frac{\chi^{2}_{tot}-\chi^{2}_{min}}{\chi^{2}_{min}/\nu},
\label{DCHI}
\end{equation}
where $\chi^{2}_{min}$ is the minimum $\chi^{2}_{tot}$ and $\nu$ is
the number of degrees of freedom.  Figure~\ref{isocontour} is a
contour plot showing the confidence region for the joint variation in
metallicity and age.  The three solid contours represent the 1-, 2-,
and 3-$\sigma$ confidence limits corresponding to $\Delta\chi^{2}=2.3,
6.14$, and $14.0$ for the two degrees of freedom, respectively.  To
further refine the best fit isochrone parameters, we ran another 40x40
metallicity and age grid at finer resolution: $-0.17\leq {\rm [Fe/H]}
\leq 0.03$ and $0.9550\leq {\rm Age (Gyr)}\leq 1.2303$.  We quantify
the 1-$\sigma$ error bounds by fitting a paraboloid to the
$\Delta\chi^2$ surface as a function of the four fit parameters ${\rm
[Fe/H]}$, Age, $(m-M)_0$, and $A_V$. The 1-$\sigma$ errors are then
the projection of the $\Delta\chi^2=1.0$ extent of this paraboloid on
the parameter axes.  We note that this procedure assumes that the
$\Delta\chi^2$ surface near the minimum is parabolic, which we find to
be a good approximation.  Also, because we minimize on $A_V$ and
$(m-M)_0$ at fixed age and metallicity, we are ignoring the additional
variance from these parameters.  Therefore, our errors are slightly
underestimated.  The resulting best fit parameters are
${\rm [Fe/H]}=-0.05\pm0.03$, ${\rm Age}=1.035\pm0.022$ Gyr,
$(m-M)_{0}=12.27\pm0.02$ mag, and A$_{V}=0.68\pm0.02$ mag.  The
best fit isochrone is overplotted on the $V,(B-V)$ and $V,(V-I)$ CMDs
in Figure~\ref{isofitcmd}, and the best fit metallicity and age are
shown as the filled square in Figure~\ref{isocontour}.

The isochrone fits reasonably well in the $V,(B-V)$ CMD, deviating
slightly too red toward the lower main sequence.  We investigate
whether an adjustment to the photometric calibration color term can
reconcile the observed main sequence and the isochrone by first
calculating the color residual between the stars and the isochrone at
fixed $V$ mag.  A line fit to the color residuals as a function of
$(B-V)$ color has a slope, $b=-0.065\pm0.007$.  This calculation
assumes that the V-band color term is correct, and only the B-band
color term needs adjusting.  The error in determining the color term,
$1-\sigma\sim0.004$ in all passbands, suggests the large, $b=-0.065$,
color term adjustment required to reconcile the observed main sequence
and theoretical isochrone does not result from an incorrect
determination of the color term alone.

The fit is poorer in the $V,(V-I)$ CMD.  Visually, the best fit
isochrone exhibits a tendency to lie redward of the main sequence in
the $V,(V-I)$ CMD.  This tendency toward the red partly results from
the broader and more asymmetric main sequence in the $V,(V-I)$ CMD.
Additionally, the overly red isochrone results from systematic errors
in the shape of the isochrone.  The isochrone fits best near
$(V-I)\sim 0.9$ and deviates redward of the main sequence toward
fainter and brighter magnitudes.  A line fit to the $(V-I)$ color
residual between the stars and isochrone at fixed $V$ mag as a
function of $(V-I)$ color does not reveal any significant trend.
Again, as with the $V,(B-V)$ CMD, the theoretical isochrones are
unable to fit the detailed shape of the cluster main sequence from the
turnoff down to the lowest observed magnitudes.  The isochrone
deviates from the observed main sequence by as much as 0.06 mag in
$(V-I)$ color at fixed $V$ mag.

\subsection{Systematic Errors}\label{syserror}
There are four main contributions to the systematic uncertainties in
the derived cluster parameters: uncertainty in $R_{V}$, photometric
calibration errors, theoretical isochrone errors, and binary star
contamination.  We address the effect of each of these uncertainties
on the derived physical parameters in turn.

We can, in principle, determine $R_{V}$ by additionally minimizing
$\chi^2_{tot}$ with respect to $R_V$.  However, in doing so we find a
best fit value of $R_{V}=2.3$ and a metallicity of ${\rm
[Fe/H]}=-0.26$.  Values of $R_{V}$ this low are rare in our galaxy
(however, see \citet{GOU01}), and metallicity measurements based on
medium resolution spectroscopic indices for NGC 1245 rule out a
metallicity this low (J.\ Marshall et al., in preparation).

We also can use infrared photometry from the 2MASS All-Sky Data
Release Point-Source Catalog
\citep{CUT03}\footnote{http://www.ipac.caltech.edu/2mass/releases/allsky/doc/explsup.html}
to place more robust constraints on $R_{V}$.  As a side note, 2MASS
photometry does not reach nearly as faint as our $BVI$ photometry.
Thus, we do not include it directly into our best fit solution for the
physical parameters of NGC 1245.  To aid in determining $R_{V}$, we
constrain the sample of stars employed in the isochrone fit to stars
with $V<16$, which corresponds to $K_{s}\sim 14.5$ mag for turnoff
stars.  At this limiting magnitude, the typical photometric error in
the 2MASS photometry is $\sigma_{K_{s}}\la 0.07$.  We transform the
2MASS $K_{s}$ passband magnitudes to the $K$-band system defined by
\citet{BES88} (the system employed in the \citet{LEJ98} color
calibration used by the Y$^{2}$ isochrones) by adding 0.04 mag to the
2MASS photometry \citep{CUT03}.  The photometric transformation of the
2MASS photometry from $K_{s}$ to $K$ does not require a color-term
\citep{CUT03}.  Using the $BVIK$ magnitudes of the $V<16$ turnoff
stars, we derive a best fit to the isochrones with $R_{V}\sim 3.2$.
However, the resulting best fit age is $\sim 1.2$ Gyr, somewhat older
than the age of $\sim 1.0$ Gyr derived from the deeper $BVI$ data
alone.  Conversely, fixing the age at $\sim 1.0$ Gyr in our isochrone
fit to the turnoff stars, we find $R_{V}\sim 3.4$ is the best fit
solution.  Figure~\ref{isofitcmd} shows the best fit isochrone to the
$BVI$ data systematically deviates from the observed main sequence at
the turnoff.  Since the 2MASS photometry limits our $BVIK$ isochrone
fit to within 1.5 mag of the turnoff, we suspect the difficulty in
determining $R_{V}$ with an age consistent with our best fit solution
arises from systematic errors in the isochrones.  We therefore adopt a
fiducial value of $R_{V}=3.2$ for our best isochrone fit but explore
the impact a systematic uncertainty in $R_{V}$ of $0.2$ has on the
cluster parameters in the following paragraph.

\begin{figure}
\plotone{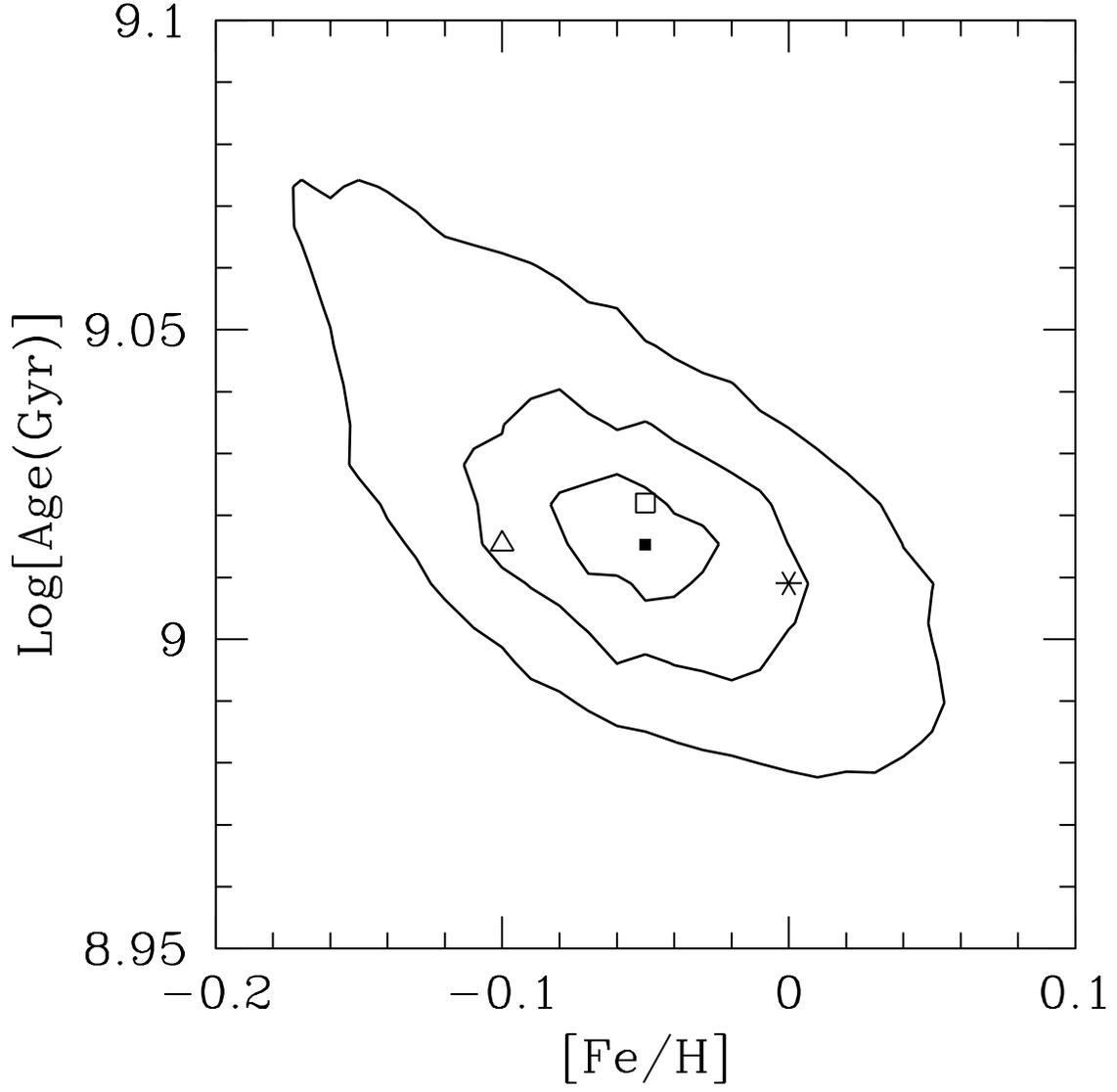}
\caption{Contour lines represent the 1-, 2-, \& 3-sigma confidence
regions on the joint variation in metallicity and age.  The filled
square shows the best fit solution.  The open square shows the best
fit solution after reducing binary star contamination.  The open
triangle shows the best fit solution assuming
a ratio of selective-to-total extinction of $R_{V}=3.0$ instead of
$R_{V}=3.2$.  The starred point shows the best fit solution assuming
an $I$-band calibration fainter by 0.02 mag.}
\label{isocontour}
\end{figure}

We quantify the effect of the systematic uncertainty in $R_V$ on the
cluster parameters by recalculating the isochrone fit assuming
$R_V=3.0$, the 1-$\sigma$ lower limit in our derived value of $R_{V}$.
The resulting best fit is ${\rm [Fe/H]}=-0.1$, Age=1.036 Gyr,
$(m-M)_{0}=12.29$ mag, and $A_{V}=0.68$ mag.  The open triangle in
Figure~\ref{isocontour} shows the resulting best fit metallicity and
age for $R_{V}=3.0$.  Since the $(B-V)$ color is more sensitive to
metallicity variations than the $(V-I)$ color, the color difference
between the main sequence in the $V,(B-V)$ CMD and $V,(V-I)$ CMD is a
good indicator of metallicity.  Unfortunately, for a fixed $A_{V}$,
variations in $R_{V}$ also alter the the relative $(B-V)$ and $(V-I)$
color of the main sequence.  This degeneracy between $R_{V}$ and
metallicity in determining the main sequence color limits our ability
to determine the metallicity.

To assess the systematic error associated with the photometric
calibration uncertainty, we refit the isochrones assuming the $I$-band
photometry is fainter by 0.02 mag.  The resulting best fit is
${\rm [Fe/H]}=0.0$, Age=1.021 Gyr, $(m-M)_{0}=12.31$, and
$A_{V}=0.62$.  The $I$-band offset solution is shown as the starred
point in Figure~\ref{isocontour}.

We attempt to investigate the systematic uncertainties in the
isochrones by fitting the Padova group isochrones \citep{GIR00} to the
CMDs.  We find that these isochrones do not match the shape of the
observed main sequence.  Systematic differences between the isochrones
and data can reach up to 0.08 mag in $(B-V)$ and $(V-I)$ colors.  The
mismatch in shape to the observed main sequence prevents precise
isochrone fits with the Padova isochrones, and we therefore cannot
reliably quantify the systematic errors in the isochrones from these
fits.  However, we find that the solar-metallicity, 1 Gyr isochrone of
the Padova group is broadly consistent with the observed data.

As an alternative method to quantify the systematic errors in the
isochrones, we calculate the physical parameters for the well studied
Hyades open cluster using an identical procedure to the NGC 1245 data.
The Hyades CMD data are selected for membership based on proper motion
and eliminated of binary star contaminants based on spectroscopic
observations (see \citealt{PIN03} for a discussion of the data and
membership selection).  Using a metallicity and age grid with
resolution $\Delta {\rm [Fe/H]}=0.01$ dex and $\Delta \log({\rm
Age})=0.006$ dex, we find best fit isochrone parameters of ${\rm
[Fe/H]}=0.10$, ${\rm Age}=670$ Myr, $(m-M)_{0}=3.23$ mag, and
$A_{V}=0.03$ mag.  The recent study by \citet{PAU03} and many previous
studies determine a value of ${\rm [Fe/H]}=0.13\pm0.01$ for the metallicity
of the Hyades using high resolution spectroscopy.  \citet{PER98} find
an age of $625\pm 50$ Myr using an independent set of isochrones.  The
Hipparcos distance to the Hyades is $(m-M)_{0}=3.33\pm 0.01$ mag
\citep{PER98}, and the extinction for the Hyades is commonly cited as
negligible.  The difference in the physical parameters of the Hyades
from our isochrone fit to the more accurate determinations of these
parameters provides the relative systematic error in our isochrone
fitting technique due to systematic errors in the isochrones.

The final systematic error source we address is the error resulting
from contamination due to unresolved binaries.  Unresolved binaries
tend to lie redward of the single-star main sequence, and this binary
``reddening'' has a larger effect on the $(V-I)$ color than the
$(B-V)$ color.  Thus, the main sequence in a $V,(V-I)$ CMD is not as
tight and shows greater intrinsic scatter than the $V,(B-V)$ CMD
because the unresolved binaries are separated in $(V-I)$ color from
the single-star main sequence more than they are in $(B-V)$ color.  An
additional selection of the main sequence in the $V,(V-I)$ CMD
provides a sample of probable cluster members with reduced
contamination by unresolved binaries.  For determining the additional
main sequence selection in the $V,(V-I)$ CMD, we trace by eye the blue
boundary of the main sequence.  The red main sequence boundary is
defined by offsetting the blue boundary by 0.08 mag in color.  A 0.15
mag color offset is needed to select most of the stars that meet the
original main sequence selection based on the $V,(B-V)$ CMD alone.
This additional main sequence selection reduces the stellar sample to
584 from the original 1004.  The best fit isochrone using stars that
meet the main sequence selection criteria in both CMDs has parameters
${\rm [Fe/H]}=-0.05$, $Age=1.052$ Gyr, $(m-M)_{0}=12.33$ mag, and
A$_{V}=0.62$ mag.  The above solution is shown as the empty square in
Figure~\ref{isocontour}.

To define the 1-$\sigma$ systematic error on the fitted parameters, we
use the difference between the best fit parameters and the parameters
determined during the above discussion of the three sources of
systematic error: uncertainty in $R_{V}$, photometric calibration and
binary star contamination.  For the 1-$\sigma$ systematic error in the
theoretical isochrones, we use the relative difference between our
isochrone fit parameters for the Hyades and the quoted values from the
literature.  To derive an overall systematic error in the cluster
parameters, these four sources of systematic error are added in
quadrature.  The resulting overall 1-$\sigma$ systematic errors are
$\sigma_{\rm [Fe/H]}=0.08$, $\sigma_{\rm Age}=0.09$ Gyr,
$\sigma_{(m-M)_{0}}=0.12$ mag, and $\sigma_{A_{V}}=0.09$ mag.  The
photometric calibration and $R_{V}$ uncertainty dominate the
systematic uncertainty in metallicity.  The systematic uncertainty in
the isochrone dominates the systematic uncertainties in the age and
distance.  Unresolved binary contamination and photometric calibration
dominate the uncertainty in the reddening.

\subsection{Comparison with Other Determinations}

The best fit metallicity we derive from isochrone fitting to NGC 1245
agrees with our independent metallicity determination using
spectroscopy of individual red-giant members (J.\ Marshall et al., in
preparation).  Medium-resolution spectroscopic indices calibrated from
high-resolution spectroscopy indicate ${\rm [Fe/H]}=-0.06\pm 0.12$,
where the error is the 1-$\sigma$ systematic error in the
spectral-index-metallicity calibration.  \citet{WEE96} measure the
metallicity of NGC 1245 using Washington photometry and obtain ${\rm
[Fe/H]}=-0.04\pm_{stat}0.05\,\pm_{syst}0.16$.  We note that NGC 1245
is commonly quoted in the literature as having a super-solar
metallicity of ${\rm [Fe/H]}=+0.14$ as given by the \citet{LYN87} open
cluster database\footnote{http://vizier.u-strasbg.fr}.  The origin of
this high metallicity for NGC 1245 is unclear; the source for the high
metallicity as given in \citet{LYN87} does not contain observations
for or even discuss NGC 1245.

Our best fit age for NGC 1245 is in agreement with the ${\rm
Age}=1.1\pm 0.1$ Gyr found by \citet{WEE96}.  \citet{CAR94} and
\citet{SUB03} find a younger age of ${\rm Age}=800$ Myr for NGC 1245.
In the case of \citet{CAR94}, the younger age most likely results from
their assumption of ${\rm [Fe/H]}=+0.14$ for the cluster.  In the case of
\citet{SUB03}, the difference in color between the main sequence turnoff
and the red-giant clump requires the younger age.  We believe the
red-giant clump of \citet{SUB03} is too red by 0.35 mag in $(B-V)$
color (as discussed in section~\ref{OBS}).  Our smaller color
difference between the main sequence turnoff and the red-giant clump
results in the older age for NGC 1245.

The main differences between this study and previous investigations of
NGC 1245 are that we find a lower overall reddening and we find no
evidence for significant differential reddening.  We find
$E(B-V)=0.21\pm 0.03$, where the error includes the systematic
uncertainty in $R_{V}$ and $A_{V}$.  Previous studies find higher
reddenings $E(B-V)=0.28\pm 0.03$ \citep{WEE96}, marginally consistent
with our determination.  The previous claims for differential
reddening across NGC 1245 have not been highly significant and even
conflicting.  \citet{CAR94} determine a north/south $(B-V)$ color
gradient of 0.04 mag arcmin$^{-1}$ with redder colors toward the
south, and \citep{WEE96} find a north/south $(B-V)$ color gradient of
0.03$\pm\,0.16$ mag arcmin$^{-1}$, except the color gradient is in the
opposite sense, with redder colors to the north.

We follow the same procedure as in \citet{CAR94} and \citet{WEE96} to
determine the differential reddening across NGC 1245 using the 1.3m
dataset.  To quantify the differential reddening across NGC 1245, we
start with the same sample of stars that were used for the isochrone
fit.  We further restrict the sample to stars with $15.0\la V\la 17.0$
mag.  The differential reddening is modeled as a linear trend
between the $(B-V)$ color of stars as a function of north/south
position.  We formally find a north/south $(B-V)$ color gradient of
0.0042$\pm\, 0.0006$ mag arcmin$^{-1}$ with redder colors to the
north.  The color gradient corresponds to a total change of $\Delta
(B-V) \sim 0.03$ over 9 arcmin.  The color gradient in the
east/west direction is half the north/south color gradient with a
similar error.  Therefore, we find a color gradient that is smaller by
an order of magnitude than previous investigations.  Furthermore,
although formally significant, we are not convinced of the color
gradient's reality because the two-dimensional aperture correction map
and PSF variations both contain errors of order 0.02 mag.

\section{Radial Profile}\label{RAD}
Stellar encounters drive an open cluster toward equipartition of
kinetic energy resulting in mass segregation of the cluster members
\citep{BIN87}.  The survey of \citet{NIL02} demonstrates that the
prominent cores of open clusters generally contain the most massive
stars, but 75\% of a cluster's mass lies in a surrounding corona
containing only low mass members of the cluster.  In this section, we
search for the signatures of mass segregation in NGC 1245 by studying
its radial surface-density profile as a function of the magnitude of
the cluster stars.

The first step in deriving the radial profile is to locate the cluster
center.  Using the stellar positions for stars with high-quality
photometry shown in the 1.3m CMD (Figure~\ref{cmds}), we spatially
smooth the number counts with a Gaussian of radius $\sim 75''$.  We do
not use the 2.4m dataset to calculate the cluster center because gaps
in the CCD array bias the derived center.  The open circle in
Figure~\ref{n1245fov} denotes the cluster center.  Our derived cluster
center is located $48''$ east and $75''$ south of the center
determined by \citet{CAR94} and $36''$ east and $36''$ south of the
center determined by \citet{SUB03}.  Since neither study gives exact
details on how the cluster center was determined, we cannot determine
the significance of these differences in the cluster center.  However,
we believe that our estimate for the cluster center is likely to be
more robust, due to our larger field of view and fainter limiting
magnitude.

\begin{figure}
\plotone{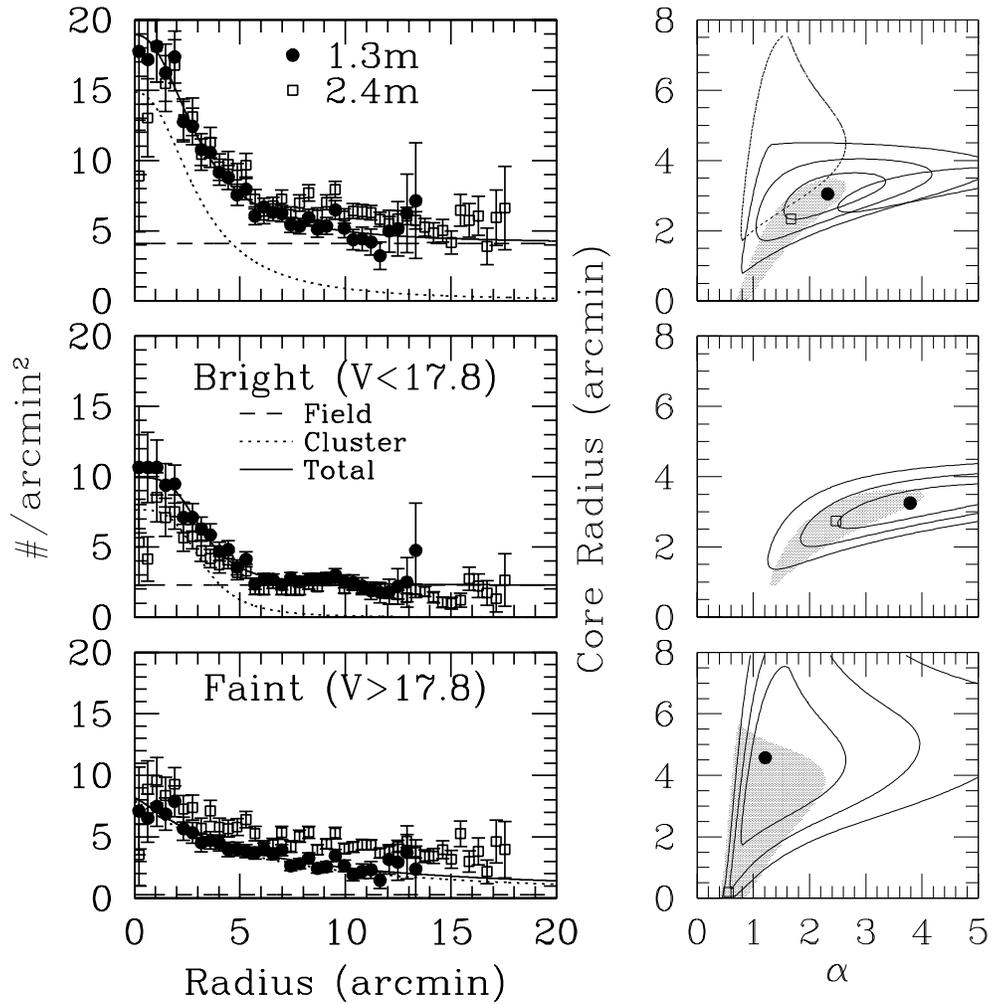}
\caption{ Left panels show the radial surface-density profiles of
stars in the NGC 1245 field for three different samples of stars.  The
top row shows the radial profile for all stars.  The middle panel
shows the profile for the subset of these stars with $V<17.8$, while
the lower panel is for $V>17.8$.  The solid circles are derived from
the 1.3m data, while the open squares are from the 2.4m data.  In each
panel, the solid line shows the best fit model to the 1.3m data, which
is composed of a cluster profile (dotted line) plus a constant
field-star surface density (dashed line).  The panels on the right
show the 68\%, 95\%, and 99\% confidence regions on the fitted
logarithmic slope of the density profile $\beta$ and core radius for
the 1.3m data.  The filled point shows the best fit model for the 1.3m
data, whereas the open square is the best fit for the 2.4m dataset.
The shaded region is the 68\% confidence region for the fit to the
2.4m dataset. In the upper-right panel, we also show the 68\%
confidence regions for the bright and faint samples.}
\label{rprof}
\end{figure}

We construct a radial surface-density profile by determining the
surface density of stars in concentric annuli of width $\sim 25''$.
The top left panel in Figure \ref{rprof} shows the surface-density
profile derived from both the 1.3m and 2.4m datasets.  We approximate
the errors as $N_i^{1/2}$, where $N_i$ is the number of stars in each
bin.  Area incompleteness sets in due to the finite size of the
detector for annuli $\ga 7.3'$ in the 1.3m dataset and $\ga 11.6'$ for
the 2.4m dataset, thereby increasing the errors.  Generally, the
surface-density profiles derived from the two datasets agree well,
except near the center of the cluster and a small overall scaling.
The scaling difference results from the fainter limiting magnitude in
the 2.4m dataset (see Figure~\ref{cmds}).  The discrepancy in the
inner annuli arises from the fact that the cluster center falls very
close to a gap in the CCD array for the 2.4m detector.  Therefore,
incompleteness affects the first few radial bins of the 2.4m
surface-density profile.  We disregard the first two radial bins for
all surface-density profiles derived from the 2.4m dataset.

As expected, the radial profile of the cluster exhibits a gradual
decline to a constant field-star surface density.  We fit the
surface-density profile to the model,
\begin{equation}
\Sigma(r) = \Sigma_f + \Sigma_0 \left[1 + \left(\frac{r}{r_c}\right)^\beta\right]^{-1},
\label{radprofeqn}
\end{equation}
where $\Sigma_f$ is the surface density of field stars, $\Sigma_0$ is
the cluster surface density at $r=0$, and $r_c$ is the core radius.
In fitting Equation~\ref{radprofeqn}, we require a positive
$\Sigma_f$.  We fit this profile to both the 1.3m and 2.4m datasets.
Although the 2.4m surface-density profile extends to larger radii, the
incompleteness near the center generally results in less
well-constrained parameters, in particular the core radius $r_c$.

Both fits are excellent: for the 1.3m dataset, $\chi^2=15.9$ for 28
degrees of freedom (dof), whereas for the 2.4m dataset, $\chi^2=39.3$
for 36 dof.  Table~\ref{tbl:table3} tabulates the best fit parameter
values and 1-$\sigma$ uncertainties for all datasets.  A parabola fit
to the envelope of $\chi^2$ values as a function of the parameter of
interest determines the parameter's error.  The solid line in the top
left panel of Figure~\ref{rprof} shows the best fit model to the 1.3m
profile.  The logarithmic slope of the surface density is $\beta=2.36
\pm 0.71$, and the core radius is $3.10 \pm 0.52$ arcmin.  The field
and central surface densities are $\Sigma_f=4.0 \pm 0.8~{\rm
arcmin}^{-2}$ and $\Sigma_0=14.7 \pm 2.9~{\rm arcmin}^{-2}$,
respectively.  At the derived distance to the cluster, $d=2850~{\rm
pc}$, the conversion from angular to physical distances is $1.207~{\rm
arcmin/pc}$.  Thus, NGC 1245 has a core radius of $\sim 2.6~{\rm pc}$.
The core radius agrees with the previous determination of $2.7 \pm
0.13$ arcmin from \citet{NIL02}.

To investigate mass segregation, we next divide the high-quality
photometry sample of stars into two subsamples: bright stars with
$V<17.8$ and faint stars with $V>17.8$.  There are roughly an equal
number of cluster stars in each subsample.  The surface-density
profiles for both the 1.3m and 2.4m data are shown in the bottom left
and middle left panels of Figure~\ref{rprof}.  We find a significant
difference between the best fit logarithmic slopes of the bright and
faint subsamples.  For the 1.3m dataset, we find $\beta=3.9 \pm 1.3$
for the bright sample, and we find $\beta=1.3\pm 0.9$ for the faint
sample.  This provides evidence, at the $\sim 2-\sigma$ level, that
the faint cluster stars in NGC 1245 are considerably more spatially
extended than the brighter stars.  The fainter stars being spatially
extended is almost certainly due to the effects of mass segregation,
which we expect to be important for a cluster of this mass and age.
We provide additional evidence for mass segregation in
Section~\ref{massfunc} and discuss the implications of this mass
segregation more thoroughly in Section~\ref{totmass}.

The radial profile of NGC 1245 implies that the faint cluster members
may actually extend significantly beyond our field of view.  The
radial profile for the 1.3m data predicts that, at 1-$\sigma$, 40 to
100\% of the faint ($V>17.8$) stars outside $10'$ of the cluster
center and within the field of view are cluster members.  The strong
cluster contribution in the outer periphery of our observed field
hampers our ability to obtain an accurate statistical subtraction of
the field star contamination necessary to determine the mass function
for NGC 1245.  We discuss this additional complication and our
solution in Section~\ref{massfunc}.

\section{Mass Function}\label{massfunc}

Typically an assumed mass-luminosity relation allows the
transformation of the observed luminosity function to derive the
present day mass function \citep{CHA03}.  In contrast, we assign a
mass to each star individually using our best fit isochrone.
Minimizing a star's individual $\chi^2_i$ (Equation~\ref{isochi})
using the best fit isochrone parameters determines the star's mass.
To determine the mass for each star, we use the high quality 2.4m
photometry shown in the bottom panel CMDs of Figure~\ref{cmds} as it
covers a much larger field of view.  We reduce field star
contamination of the cluster mass function by restricting the sample
to stars with $\chi^{2}_{i}<0.04$.  This $\chi^{2}_{i}$ cutoff selects
a region in the $V,(B-V)$ CMD roughly similar to the main sequence
selection used in the isochrone fit but naturally selects stars based
on their proximity to the isochrone in all three passbands.

Despite the large field of view of the 2.4m data, the radial profile
of the cluster suggests that there exists a significant number of
low mass cluster members out to the limits of our field of view.
Lacking observations of a field offset from the cluster, the
Besan\c{c}on theoretical galaxy model \citep{ROB86} provides an
alternative method to correct the observed mass function for field
star contamination.  We obtain from the Besan\c{c}on galaxy model
electronic database the stellar properties and photometry for a
one-square-degree field centered on the cluster's galactic coordinates
without observational errors.  A spline fit to the median photometric
error in 0.5 mag bins in each passband allows us to simulate the
observed errors in the theoretical galaxy field photometry.  We add a
zero-mean, unit-standard-deviation Gaussian random deviate scaled to
the observed photometric error model to the theoretical galaxy field
photometry.  We apply the same error selection criteria
($\sigma<0.08$) as was applied to the observed CMD and place a
saturation cutoff at V=14.0 mag to the theoretical CMD.

\begin{figure}
\plotone{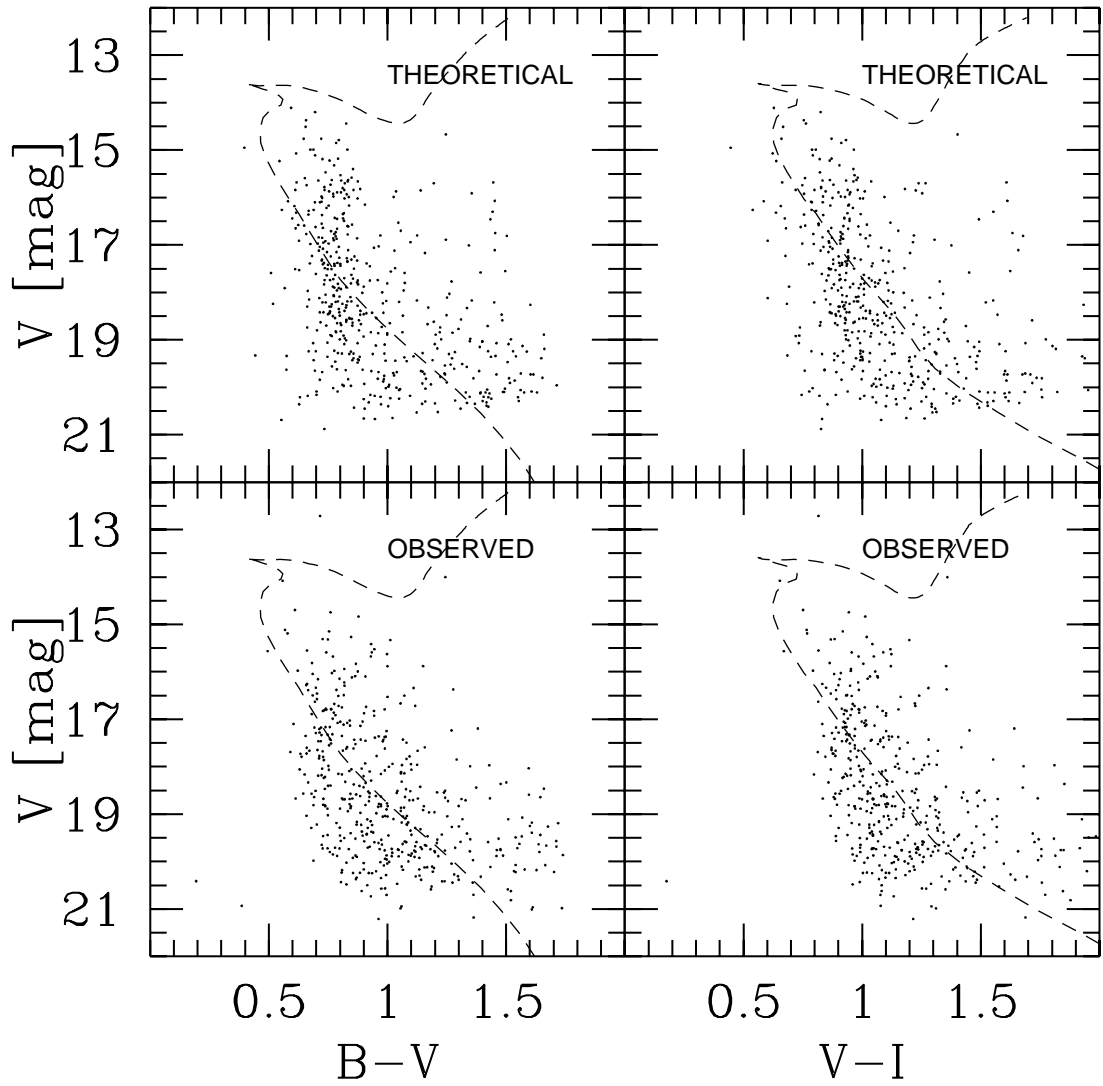}
\caption{Top Panels: Besan\c{c}on theoretical galaxy model $V,(B-V)$
and $V,(V-I)$ CMDs for the Galaxy position of NGC 1245 \citep{ROB86}.
Photometric errors and selection criteria are included in the
theoretical galaxy to match the observed CMDs shown in the bottom
panels.  Bottom Panels: Observed $V,(B-V)$ and $V,(V-I)$ CMDs for
stars beyond 12.7 arcmin of the cluster center using the 2.4m data.
In all panels, the dashed line shows the best fit
isochrone.}
\label{fieldcmd}
\end{figure}

To make a qualitative comparison between the theoretical field CMD and
the observed CMD, we adjust the theoretical CMD for the smaller
observed 2.4m field of view.  In the comparison, a star is included in
the theoretical CMD only if a uniform random deviate is less than the
ratio between the two field of views.  Figure~\ref{fieldcmd} compares
the CMDs of the theoretical galaxy model in the top panels and the
observations for stars beyond 12.7 arcmin of the cluster center for
the 2.4m data in the bottom panels.  The dashed line shows the best
fit isochrone for reference.  As shown in Figure~\ref{fieldcmd}, the
theoretical and observed field CMDs are qualitatively very similar.

\begin{figure}
\plotone{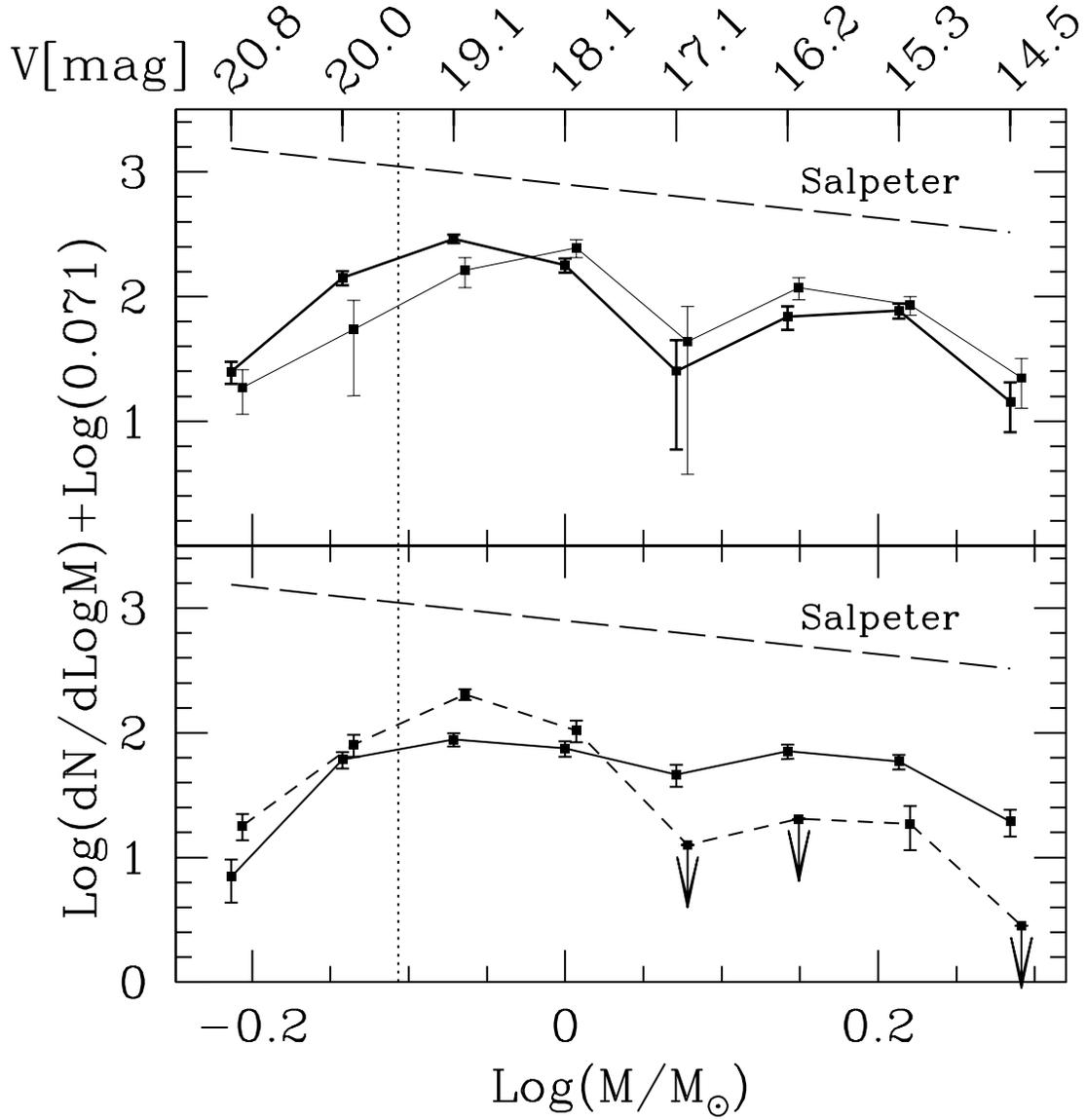}
\caption{Top Panel: Mass function using a theoretical galaxy model
(heavy solid line) and outer periphery of the field of view (light solid line)
for determining the field contamination.  The vertical line delineates
mass bins complete to better than 90\%.  Bottom Panel: Mass function
for stars inside (solid line) and outside (dashed line) 4.3 arcmin.
The down pointing arrows represent mass bins with negative values, and
the point on the tail represents the 2-sigma upper
limit.}
\label{massfuncfig}
\end{figure}

To calculate the field star contamination in the mass function, we
first obtain the mass for the stars in the theoretical field CMD and
apply the same $\chi^{2}_{i}<0.04$ selection in the same manner as the
observed field CMD.  For both the theoretical field CMD and observed
cluster CMD, we calculate mass functions using even log-spaced
intervals of 0.071 $\log(M_{\odot})$.  Number counts from the entire
one-square-degree field of view provide the basis for the theoretical
galaxy mass function.  The larger field of view of the theoretical
galaxy model than the observed field requires scaling of the number
counts and corresponding Poisson error of the theoretical galaxy mass
function by the ratio of field of views.  Subtracting the scaled
theoretical galaxy field number counts from the observed number counts
provides the cluster mass function.  The cluster mass function is
shown as the heavy solid line in the upper panel of
Figure~\ref{massfuncfig}.  The top axis shows the median apparent V
magnitude for the corresponding mass bin.  The vertical dashed line
delineates mass bins that are complete to better than 90\%.  Comparing
the cluster mass function to a mass function calculated by subtracting
the theoretical galaxy number counts without applying the photometric
error selection criteria determines the completeness of a mass bin.
The long-dashed line labeled `Salpeter' in Figure~\ref{massfuncfig}
illustrates the typical observed mass function slope, $\alpha=-1.35$.
A power law in linear-mass, linear-number-count space fit with weights
to the mass function bins not affected by incompleteness yields a
slope steeper than Salpeter, $\alpha=-3.12\pm 0.27$.

To verify the validity of using the theoretical galaxy field in place
of an observed off-cluster field, we calculate an alternative mass
function using the observed field outside 12.7 arcmin of the cluster
center for field contamination subtraction.  The entire 2.4m field of
view again provides the basis for the observed number counts.  The
smaller field of view of the outer periphery requires scaling the
field contamination number counts and corresponding Poisson error by
the relative area factor, $f=6.4$.  The light-solid line in the upper
panel of Figure~\ref{massfuncfig} shows the resulting cluster mass
function using the outer periphery for determining the field
contamination.  The mass bin positions are offset slightly for
clarity.  The two mass functions agree within $1-\sigma$ for all bins
with $\log{M}>0$ but disagree significantly at the low mass end.  The
theoretical galaxy subtracted mass function contains greater number
counts at the low mass end, consistent with our conclusion that the
outer regions of our observed field of view contain a significant
number of low mass cluster stars.  Both mass functions contain a drop
in the number counts at $\log(M)=0.071$.  We question the significance
of this drop in the number counts.  The drop occurs at a magnitude
with some of the heaviest field star contamination, making the field
star contamination subtraction less certain.

To further the evidence for mass segregation in NGC 1245, we calculate
the mass function inside and outside a radius of 4.2 arcmin from the
cluster center.  According to the cluster radial profile, this radius
roughly separates half of the cluster members.  The inner and outer
mass functions are shown as the solid and short-dashed lines,
respectively, in the lower panel of Figure~\ref{massfuncfig}.  The fit
to the inner region mass function yields a shallower slope that is
consistent with Salpeter, $\alpha=-1.37\pm\,0.2$.  Ignoring the
highest mass bin, the best fit slope for the inner region is even
shallower, $\alpha=-0.56\pm\,0.28$.  We find that the outer region is
highly enriched in low mass stars.  The outer region mass function has
a very steep slope, $\alpha=-7.1\pm\, 1.2$.  The mass bins with the
down pointing arrows have negative values, and the point represents
the 2-$\sigma$ upper confidence limit.  Since the mass function is fit
in linear space, the fit includes bins with negative values.

The radial profile and mass function demonstrate that a significant
fraction of the low mass cluster members reside in the outer periphery
of the cluster.  Previous studies of NGC 1245 generally find Salpeter
and shallower slopes for the mass function \citep{CAR94,SUB03}.  As
warned in \citet{SUB03}, the smaller field of view of these previous
studies from which to determine the field contamination biases the
mass function against low mass members that reside in the outer
periphery of the cluster.  The mass functions from the previous
studies agree well with our mass function for the interior of NGC
1245, but we conclude that a steeper slope than Salpeter,
$\alpha=-3.12\pm 0.27$, is more appropriate for NGC 1245 as a whole
down to the completeness limit of $M=0.8 M_{\odot}$.

\begin{figure}
\plotone{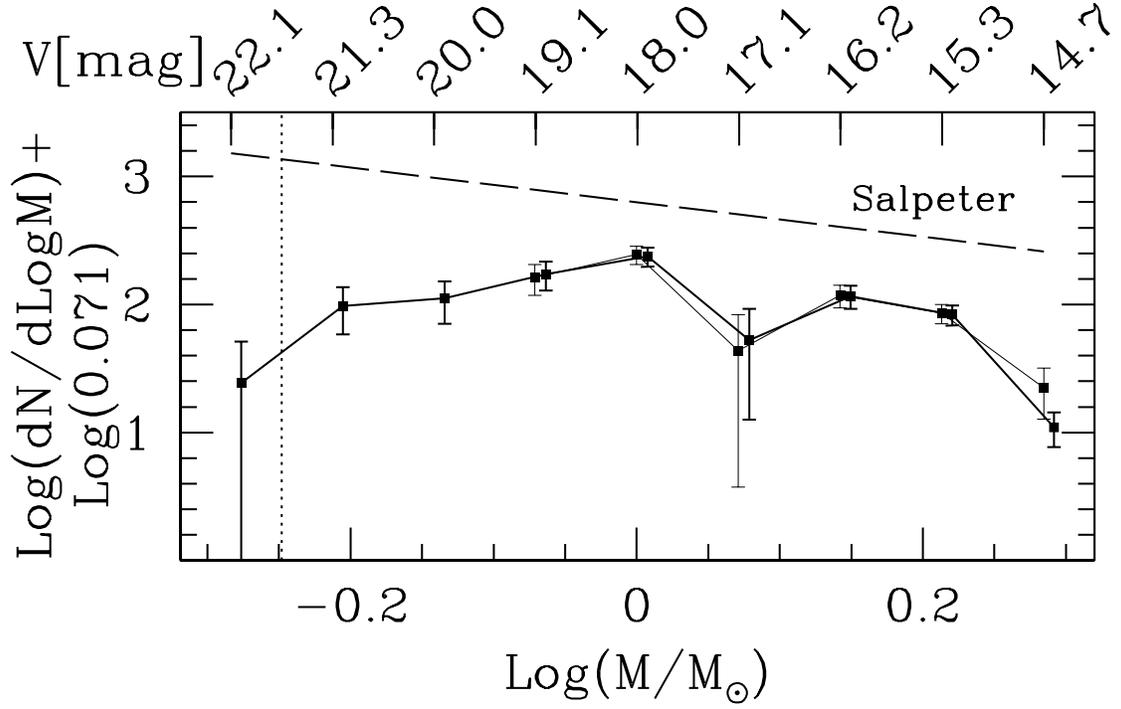}
\caption{Heavy solid line shows the mass function calculated from the
  deep 2.4m data (Figure~\ref{cmds2}) using the outer periphery of the
  field of view for determining the field contamination.  The vertical
  line delineates mass bins complete to better than 90\%.  The thin
  solid line reproduces the equivalent mass function based on the
  shallow 2.4m data as shown as the dashed line in the upper panel of
  Figure~\ref{massfuncfig}.}
\label{massfuncfig2}
\end{figure}

Our deeper CMD obtained from combining the MDM 2.4m data
(Figure~\ref{cmds2}) allows us to extend the mass function to even
lower masses.  Determining the mass function to lower masses improves
our mass estimate of the cluster, calculated in the following section.
To calculate the mass function, we follow the same procedure as
outlined above.  At these fainter magnitudes, the theoretical galaxy
field overpredicts the number of stars. Thus we must resort to using
the outer periphery of the MDM 2.4m field of view for estimating the
field contamination.  The heavy solid line in
Figure~\ref{massfuncfig2} shows the resulting mass function based on
the deeper CMD.  The light solid line reproduces the equivalent mass
function from Figure~\ref{massfuncfig} based on the shallower CMD
data.  The vertical dashed line represents the completeness limit at
$M=0.56 M_{\odot}$.  The deep mass function has a slope,
$\alpha=-0.5\pm 0.4$, significantly steeper than the slope for the
shallow mass function, $\alpha=-2.8\pm 0.8$.  The significantly
differing slopes suggest there is a turnover in the mass function at
$\log(M/M_{\odot})=-0.05$.  Fitting to the last three complete mass
bins in the deep mass function yields a slope, $\alpha=1.8\pm 0.5$.
The low mass end of the deep mass function likely does not fall as
steeply as measured, since we use the outer periphery of the MDM 2.4m
field of view to quantify the field contamination, which likely
contains many low mass cluster members.  Thus, the measured slope at
the low mass end represents an upper limit.

\section{Total Mass and Dynamics}\label{totmass}
Numerous dynamical processes take place in open clusters.
Interactions between the cluster members dramatically affect binary
and planetary systems, mass segregation within the cluster, and the
timescale for cluster dissolution \citep{HUR02A,HUR02B,GIE97,POR01}.
The cluster's initial total mass sets the timescale for these
dynamical processes and governs its dynamical evolution.
Unfortunately, several difficulties prevent a completely empirical
determination of the total mass of NGC 1245.  First, as we show in
Section~\ref{RAD}, the surface-density profile suggests the cluster
extends significantly beyond our field of view.  Second, a substantial
population of low mass cluster members exists below the completeness
limit of our mass function.  Third, we do not resolve binary systems.
Therefore, the total mass in {\it observed} cluster members ($\sim 820
M_\odot$) only places a lower limit to the current cluster mass.
Finally, even overcoming the heretofore difficulties, stellar
evolution and evaporation results in additional complications for
determining the initial birth mass of the cluster.

In order to provide a rough estimate of the current total cluster
mass, we must adopt a number of assumptions in order to extrapolate
the observed mass to the total mass.  We must account for stars
outside our survey area and stars fainter than our magnitude limit.
The total cluster mass determined by integrating the radial profile
model we fit in Section~\ref{RAD} slowly converges and is
unrealistically large for an integration to infinity.  In reality, the
tidal field of the Galaxy truncates the outer radius of the cluster.
Therefore, we refit the observed dataset to the King profile
\citep{KIN62},
\begin{equation}
\Sigma = \Sigma_f + \Sigma_0 \left\{ \left[ 1 +
\left(\frac{r}{r_c}\right)^2 \right]^{-1/2} - \left[ 1 +
\left(\frac{r_t}{r_c}\right)^2 \right]^{-1/2} \right\}^{2},
\label{king}
\end{equation}
where $r_t$ is the tidal radius of the cluster.  Note that, as $r_t
\rightarrow \infty$, this profile reduces to the radial profile given
in Equation~\ref{radprofeqn} for $\beta=2$.  For a given set of
parameters; $\Sigma_0$, $r_c$, and $r_t$; the integral over
Equation~\ref{king} determines the total number of stars,
\begin{equation}
N=\pi r_c^2 \Sigma_0 \left\{ \ln(1+x_t) -{[3(1+x_t)^{1/2}-1][(1+x_t)^{1/2}-1]}\over{1+x_t}\right\},
\label{ntot}
\end{equation}
where $x_t\equiv (r_t/r_c)^2$ \citep{KIN62}.  The total mass is then $M=\langle m\rangle_{obs} N$, where $\langle m\rangle_{obs} =1.00M_\odot$ is the average mass of observed cluster stars.

Unfortunately, King profile fits to either dataset yield only lower
limits for the value of $r_t$.  The lower limit only for $r_{t}$ is
primarily due to the covariance between $r_t$ and $\Sigma_f$.
Therefore, we need an additional constraint on either the surface
density of field stars or the tidal radius in order to determine the
total cluster mass.  Attempting to constrain the fit using values of
$\Sigma_f$ determined from the theoretical galaxy field star counts
(see Section~\ref{massfunc}) results in unrealistically large values
of $r_{t}>40~{\rm arcmin}$.  Thus, we adopt a different approach.

The tidal radius of the cluster is set by both the local tidal field
of the Galaxy and the total mass of the cluster.  We can estimate the
tidal radius as the radius of the Lagrange point \citep{BIN87},
\begin{equation}
r_t=\left(\frac{M}{3M_{\rm MW}}\right)^{1/3} D,
\label{rtidal}
\end{equation}
where $D$ is the distance of the cluster from the center of the
Galaxy, $M$ is the total mass of the cluster, and $M_{MW}$ is the mass
of the Galaxy interior to $D$.  For our derived distance of
$d=2850~{\rm pc}$, assuming $R_0=8~{\rm kpc}$, and Galactic
coordinates ($l=147{^\circ\hskip-2.4pt}.6,
b=-8{^\circ\hskip-2.4pt}.9$), the Galactocentric distance of NGC 1245
is $D=10.50~{\rm kpc}$.  For a flat rotation curve with $v=220~{\rm
km~s^{-1}}$, we find $M_{\rm MW}=1.18 \times 10^{11} M_\odot$, and
thus $r_t \simeq 20~{\rm pc} (M/2500M_\odot)^{1/3}$.

We use equations (\ref{ntot}) and (\ref{rtidal}) to solve for the
total cluster mass $M$ and tidal radius $r_t$ while simultaneously
constraining $r_c, \Sigma_f$, and $\Sigma_0$ from fitting
Equation~\ref{king} to the observed radial surface-density profile.
However, we must first account for low mass stars that are below our
detection limit.  To do this, we use the best fit mass
function (${\rm d}N/{\rm d}\log M \propto M^{\alpha}$, with
$\alpha=-3.12$) to the turnover in the mass function at $0.85M_\odot$.  For
stars between the hydrogen-burning limit and $0.85M_\odot$, we assume
$\alpha=1.0$.  Thus, our assumed cluster mass function is
\begin{equation}
\label{eqn:massfunc}
\frac{ {\rm d}N}{{\rm d}\log M}
\propto
\left\{
\begin{array}{ll}
M^{-3.12}, &  \quad \mbox{$M>0.85M_\odot$},\\ 
M^{1.0},   &  \quad \mbox{$0.08M_\odot<M<0.85M_\odot$}
\end{array}
\right.
\end{equation}
The adopted mass function slope at the low mass end, $\alpha=1.0$ is a
compromise between our measured lower limit to the slope, $\alpha=1.8$
(see Section~\ref{massfunc}), and other clusters that have an observed
mass function slope as shallow as $\alpha\sim 0.4$ at the low mass end
\citep{PRI01,BOU03}.  This yields a total cluster mass (in our field
of view) of $\sim 1312 M_\odot$ and a correction factor of $1.61$ to
the observed mass in cluster stars.

We use an iterative procedure to determine $M$ and $r_t$.  We assume a
value of $r_t$ and then fit the King profile (Equation~\ref{king}) to
the observed radial density profile.  The best fit King profile yields
trial values of $r_c, \Sigma_f, \Sigma_0$, and the total number of
cluster stars via Equation~\ref{ntot}.  We then apply the correction
factor to account for stars below our completeness limit to determine
the total cluster mass, $M=1.61\langle m\rangle_{obs} N$.  We then use
Equation~\ref{rtidal} to predict a new value for $r_t$.  The procedure
iterates until convergence with a tidal radius of $r_t\simeq 20~{\rm
arcmin}$ ($16.5~{\rm pc}$) and a total mass of $M=1312\pm 90 M_\odot$,
where the error is the 1-$\sigma$ statistical error from the fit.  We
determined the total cluster mass using the shallow 2.4m dataset as
this is the dataset used to determine the mass correction factor and
surface-density profile.

The mass correction factor, sensitive to the mass function slope at
the low mass end, dominates the uncertainty in the cluster mass.  For
example, varying the logarithmic slope of the mass function below
$0.85M_\odot$ in the range $0.6\le \alpha\le 1.4$ changes the derived total
mass by $+200 -140 M_\odot$. 

Our derived mass does not account for the presence of unresolved
binaries.  Without knowledge of the properties of the binary
population, it is difficult to assess the magnitude of the correction.
However, for a large binary fraction, the correction could be as large
as 50\%.  We have also not attempted to correct the derived
present-day mass for evolutionary effects, such as evaporation or
stellar evolution, to derive an initial cluster mass.  For our adopted
mass function (Equation~\ref{eqn:massfunc}), the correction to the
total cluster mass due to stellar evolution is small, $\sim 12\%$.
About $10\%$ of the cluster stars are lost to the Galaxy per
relaxation time \citep{SPI87, POR01}.  Since the current age of NGC
1245 is approximately 8 times larger than the current relaxation time
(see below), it is possible that the cluster has lost an appreciable
amount of mass over its lifetime.  The numerical N-body simulations of
\citet{POR01} predict the initial cluster mass based on the current
number of red-giant stars.  Using their Equation~1, an age of 1 Gyr
for NGC 1245 and a total of 40 observed red-giant stars yields an
initial total cluster mass, $M_{o}=3120 M_{\odot}$.

The relevant quantity that determines the timescale for dynamical
evolution of a stellar cluster is the half-mass relaxation timescale
\citep{SPI87},
\begin{equation}
t_{rlx}=0.138 \left(\frac{r_{hm}^3}{G \langle m\rangle}\right)^{1/2} \frac{N^{1/2}}{\ln\Lambda}.
\end{equation}
Here $r_{hm}$ is the radius enclosing half of the total mass of the
cluster, $\langle m \rangle \simeq 0.63M_\odot$ is the average mass of
cluster members, and $\ln \Lambda \simeq \ln 0.4 N$ is the Coulomb
logarithm.  This is roughly the average time over which the velocity
of a typical cluster member changes by order unity due to random
encounters with other cluster members.  For NGC 1245, $r_{hm}=3.8~{\rm
pc}$, and $t_{rlx}\simeq 130~{\rm Myr}$.  Since the age of NGC 1245 is
$\sim 1~{\rm Gyr}$, we expect this cluster to be dynamically quite
evolved.  Therefore, the evidence of mass segregation found in
Sections \ref{RAD} and \ref{massfunc} is not surprising.

\section{Conclusion}\label{CON}

In this paper, we report our results based on $BVI$ photometry of the
open cluster NGC 1245 using the MDM 1.3m and 2.4m telescopes.  Here,
we improve on the observations for this cluster by covering a
six-times greater area, obtaining the first CCD $I$-band photometric
data, and acquiring the first CCD data under photometric conditions.
With these significant improvements over earlier studies, we are able
to determine more precisely the physical parameters of the cluster as
a whole, rather than just the inner cluster core.  Our photometric
data resolve some of the confusion in the literature regarding the
correct photometric zeropoint for this cluster \citep{WEE96,SUB03}.
Based on isochrone fits employing the Y$^{2}$ calculations
\citep{YI01}, we confirm the findings of \citet{WEE96} and (J.\
Marshall et al., in preparation) that this cluster has a slightly
subsolar metallicity, ${\rm [Fe/H]}=-0.05\pm 0.03~{\rm
(statistical)}\pm 0.08~{\rm (systematic)}$.  Our best fit age is
$1.04\pm 0.02\pm 0.09$ Gyr.  In contrast to previous studies, we do
not find evidence for significant differential reddening.  We find a
$V$-band extinction of $A_{V}=0.68\pm 0.02\pm 0.09$, and with the aid
of 2MASS $K_{s}$ photometry, we constrain the ratio of absolute to
selective extinction to be $R_{V}=3.2\pm0.2$.  The resulting absolute
distance modulus is $(m-M)_{0}=12.27\pm 0.02\pm 0.12$.

With the large field of view provided by the MDM 8K Mosaic imager, we
find NGC 1245 is a highly relaxed cluster; a majority of the low mass
cluster members reside in an extended halo.  The mass function slope,
$\alpha=-3.12\pm 0.27$, down to $M=0.85 M_{\odot}$ is steeper than the
Salpeter value of $\alpha=-1.35$, found by earlier studies.  The
previous studies did not have a sufficient field of view to detect the
low mass cluster members that preferentially reside in the outer
periphery of the cluster.  Based on the observed stellar
surface-density profile and an extrapolated mass function, we derive a
total cluster mass, $M=1300\pm 90\pm 170 M_{\odot}$.

This paper is the first in a series for the Search for Transiting
Extrasolar Planets in Stellar Systems (STEPSS) project.  The main goal
of the STEPSS project is to determine the fraction of stars with
close-in, Jupiter-sized planets using the photometric transit
technique.  By searching for extrasolar planets in open clusters of
known metallicity, stellar density, and age, we hope to constrain the
impact of these parameters on the resulting planetary systems.  The
accurate physical parameters for NGC 1245 derived in this paper will
be used in the analysis of the 19 night run targeting NGC 1245 to
search for transiting extrasolar planets (B.\ S.\ Gaudi et al., in
preparation).  The NGC 1245 transit search, consisting of 5 minute
time resolution $I$-band photometry taken over 19 nights, will also
provide valuable information on the variable star content of NGC 1245
(J.\ Pepper et al., in preparation).

\acknowledgements This publication makes use of data products from the
Two Micron All Sky Survey, which is a joint project of the University
of Massachusetts and the Infrared Processing and Analysis
Center/California Institute of Technology, funded by the National
Aeronautics and Space Administration and the National Science
Foundation.  This work was supported in part by NASA through a Hubble
Fellowship grant from the Space Telescope Science Institute, which is
operated by the Association of Universities for Research in Astronomy,
Inc., under NASA contract NAS5-26555; the Menzel Fellowship from
the Harvard College Observatory; and by NASA grant NAG5-13129.

\begin{deluxetable}{rccccl}
\tablecaption{MDM 2.4m Observations\label{obsdat24}}
\tablehead{\colhead{Date (2001)} & \colhead{\#Exps} & \colhead{Filter} & \colhead{Exp (s)} & \colhead{FWHM (arcsec)} & \colhead{Comments}}
\startdata
Oct. 27 & 9 & V & 300 & 1.4 & overcast  \\
\nodata & 9 & I & 300 & 1.1 & \nodata   \\
Nov.  7 & 7 & V & 300 & 1.8 & partly cloudy \\
\nodata & 7 & I & 300 & 1.2 & \nodata   \\
      8 & 5 & B & 240 & 1.6 & cirrus    \\
\nodata & 5 & V & 240 & 1.5 & \nodata   \\
\nodata & 5 & I & 240 & 1.3 & \nodata   \\
     10 & 5 & B & 240 & 1.4 & cirrus    \\
\nodata & 5 & V & 240 & 1.6 & \nodata   \\
\nodata & 5 & I & 240 & 1.3 & \nodata   \\
     11 & 5 & B & 120 & 1.1 & clear   \\
\nodata & 5 & V & 120 & 1.1 & \nodata \\
\nodata & 5 & I & 120 & 1.0 & \nodata \\
\nodata & 1 & I & 240 & 0.9 & \nodata \\
\enddata
\end{deluxetable}

\begin{deluxetable}{rccccl}
\tablecaption{MDM 1.3m Observations\label{obsdat13}}
\tablehead{\colhead{Date (2002)} & \colhead{\#Exps} & \colhead{Filter} & \colhead{Exp (s)} & \colhead{FWHM (arcsec)} & \colhead{Comments}}
\startdata
Jan. 31 & 2 & B & 120 & 2.0 & non-photometric  \\
\nodata & 3 & V &  60 & 1.9 & \nodata \\
\nodata & 3 & I &  30 & 1.8 & \nodata \\
\nodata & 3 & I &  20 & 1.8 & \nodata \\
Feb.  5 & 6 & B & 120 & 1.7 & non-photometric  \\
\nodata & 4 & V &  60 & 1.6 & \nodata \\
\nodata & 6 & I &  20 & 2.0 & \nodata \\
Feb.  7 & 2 & B &  90 & 1.6 & photometric \\
\nodata & 2 & V &  60 & 1.4 & \nodata \\
\nodata & 3 & I &  20 & 1.7 & \nodata \\
\enddata
\end{deluxetable}

\begin{deluxetable}{ccccccc}
\tablewidth{0pt}
\tablecaption{Fitted Parameters and $1-\sigma$ Errors to Surface-Density Profiles.\label{tbl:table3}}
\tablehead{
\colhead{Dataset} & \colhead{Sample} & \colhead{$\Sigma_f$} & \colhead{$\Sigma_0$} & \colhead{$r_c$} & \colhead{$\beta$} & \colhead{$\Sigma_f$ (model)}\\
\colhead{ } & \colhead{ } & \colhead{$\#/{\rm arcmin}^{2}$} & \colhead{$\#/{\rm arcmin}^{2}$} & \colhead{arcmin} & \colhead{$\#/{\rm arcmin}^{2}$} & \colhead{ } \\
\colhead{ } & \colhead{ } & \colhead{($\#/{\rm pc}^{2}$)} & \colhead{($\#/{\rm pc}^{2}$)} & \colhead{(pc)} & \colhead{ } & \colhead{($\#/{\rm pc}^{2}$)}
}
\startdata
1.3m & All & $4.05\pm 0.84$ & $14.72 \pm 2.88$ & $3.10 \pm 0.52$ & $2.36\pm 0.71$ & $5.087\pm 0.040$\\
 & &($5.90 \pm 1.22$) & ($21.43\pm 4.20$) & ($2.57\pm 0.43$) & &($7.401\pm 0.058$)\\
1.3m & Bright & $2.29\pm0.23$ & $7.61\pm1.52$ & $3.27\pm0.51$ & $3.87\pm1.34$ & $2.189\pm 0.024$\\
 & &($3.33\pm0.34$) & ($11.08\pm 2.21$) & ($2.72\pm 0.42$) & &($3.187\pm 0.035$)\\
1.3m & Faint & $0.49\pm 1.93$ & $7.76\pm 3.47$ & $4.61\pm2.42$ & $1.29\pm0.88$ & $2.887 \pm 0.028$\\
 & &($0.71\pm2.80$) & ($11.29\pm5.05$) & ($3.82\pm2.00$) & &($4.203\pm0.041$)\\
2.4m & All & $4.70\pm1.10$ & $18.89\pm12.19$ & $2.22\pm1.44$ & $1.66\pm0.78$ & $5.440\pm0.039$\\
 & &($6.84\pm1.61$) & ($27.51\pm17.8$) & ($1.84\pm1.19$) & &($7.920\pm 0.057$)\\
2.4m & Bright & $1.59\pm0.28$ & $7.58\pm3.03$ & $2.75\pm0.95$ & $2.53\pm1.06$ & $2.053\pm0.025$\\
 & &($2.32\pm0.40$) & ($11.04\pm4.41$) & ($2.28\pm0.79$) & &($2.989\pm0.036$)\\
2.4m & Faint & $1.54\pm3.14$ & $29.17\pm 43.21$ & $0.40\pm5.01$ & $0.70\pm1.29$ & $3.379\pm0.030$\\
 & &($2.24\pm4.57$) & ($42.46\pm62.91$) & ($0.33\pm4.15$) & &($4.919\pm 0.044$)\\
\enddata
\end{deluxetable}

\end{document}